\documentclass[aip,jcp,twocolumn,showpacs]{revtex4}

\usepackage{amssymb,amsmath,bm,latexsym}
\usepackage{epsfig}
\usepackage{color}
\def\Fbox#1{\vskip1ex\hbox to 8.5cm{\hfil\fboxsep0.3cm\fbox{%
  \parbox{8.0cm}{#1}}\hfil}\vskip1ex\noindent}  %%  {TEXT} in BOX

\begin{document}

\title{
Distribution of Diffusion Constants and Stokes-Einstein Violation
in supercooled liquids 
} 

\author{
Shiladitya Sengupta$^{1,2}$ and Smarajit Karmakar$^{2}$
}

\affiliation{
$^{1}$Theoretical Sciences Unit, Jawaharlal Nehru Centre for
Advanced Scientific Research, Jakkur Campus, Bangalore 560 064,
India.\\  
$^{2}$TIFR Centre for Interdisciplinary Sciences, 21 Brundavan
Colony, Narsingi,Hyderabad 500075, India.}  

\pacs{64.70.P-}{Glass transitions}
\pacs{64.70.Q-}{Theory and modeling of Glass transitions}
\pacs{61.43.Fs}{Glasses Structure}

\begin{abstract}
It is widely believed that the breakdown of the Stokes-Einstein
(SE) relation between the translational diffusivity and the shear
viscosity in supercooled liquids is due to the development of
dynamic heterogeneity {\it i.e.} the presence of both slow and
fast moving particles in the system. In this study we
\emph{directly} calculate the distribution of the diffusivity for
a model system for different temperatures in the supercooled
regime. We find that  with decreasing temperature, the
distribution evolves from Gaussian to bimodal indicating that on
the time scale of the typical relaxation time, mobile (fluid
like) and less mobile (solid like) particles in the system can be
\emph{unambiguously} identified. We also show that less mobile
particles obey the Stokes-Einstein relation even in the
supercooled regime and it is the mobile particles which show
strong violation of the Stokes-Einstein relation in agreement with
the previous studies on different model glass forming
systems. Motivated by some of the recent studies where an ideal 
glass transition is proposed by randomly pinning some fraction 
of particles, we then studied the SE breakdown as a function of 
random pinning concentration in our model system. We showed that
degree of SE breakdown increases quite dramatically with increasing
pinning concentration, thereby providing a new way
to unravel the puzzles of SE violation in supercooled liquids in
greater details. 
\end{abstract}

\maketitle

\section{Introduction}\label{sec:intro}

Dynamic heterogeneity (DH), which simply means multiple time scale
relaxation processes in the system, is ubiquitous in all glass
forming liquids and even in granular materials and gels. In many
proposed theories of glass transition it is assumed without any
explicit proof that there exists a broad distribution of
relaxation times to explain many anomalous behaviours observed in
supercooled liquids, {\it e.g.} breakdown of the Stokes-Einstein
(SE) relation. In normal liquids, the shear viscosity is related
to the translational diffusivity of a probe particle {\it via} the
Stokes-Einstein relation \cite{SER08HM,SER05Einstein,SER87LL}
(discussed later). It has been shown extensively \cite{Th93HS,
  SEB94KA, SEB95TK, SEB97CS, SEB00EdigerReview, SEB01MFCR,
  SEB01ML, 06KSD, SEB06Chen, SEB06BPS, SEB09Xu, SEB10M,
  pap:LaNave-etal, 12SKDS} that when a liquid is supercooled, the
measured self diffusivity becomes much larger than the value predicted
by the SE relation. The shear viscosity is often substituted by the
relaxation time which is roughly proportional to each other in the
relevant temperature range (See ref.\cite{shiJCP2013} for an
in-depth discussion on this issue). Phenomenological arguments
\cite{Th93HS, SEB95TK} considering supercooled liquids to consist
of mobile ``fluid-like'' and less mobile ``solid-like'' regions,
can explain naturally the decoupling between the translational
diffusion and the relaxation time. The average diffusivity is
predominantly determined by the ``fluid like'' regions whereas the
average relaxation time is dominated by the ``solid-like''
regions. 

The existence of transient clusters of mobile and less mobile
particles has been directly shown in many different studies
\cite{DH-direct,marcoMazza,09KDS,10KDSa}. In \cite{06KSD},
distributions of displacements of particles were calculated for
different densities for hard sphere fluid at different time
intervals. It was shown that at large densities close to the glass
transition density in this model, the distributions of
displacements of particles show appearance of bimodality. A
suitable cut off was then chosen to define the slow and fast
particles in the system. They were also able to show that slow
particles were largely obeying the SE relation over the whole
density range studied and it is the fast particles which strongly
violates the SE relation.

However, sometimes, {\it e.g.}, in the phenomenological theories of
SE breakdown, it is more natural to describe DH in terms of a
distribution of diffusivity and relaxation times rather than
displacements of particles. There are \emph{indirect} evidences
which support the existence of such a distribution - for example,
the universal exponential tail in the Van Hove functions seen
for many supercooled liquids \cite{07CBK}. Consider an extreme
case where the system has regions with two diffusivity - one for
``solid like'' ($D_1$ ) and the other for ``fluid like'' regions
($D_2$). Hence a distribution of diffusivity can be written as
$p(D, t) = A\delta(D - D_1) + B\delta(D - D_2)$ where $A$ and $B$ are
fixed by the normalization condition and the amount of solid like
and fluid like regions. Now if we calculate the van Hove
correlation function as  

\begin{equation}
G_s(x,t) = \int \, dD \, p(D,t) \, g(x|D,t),
\label{vanHoveEq}
\end{equation}
where $g(x|D,t) = \frac{1}{\sqrt{4\pi D
    t}}\exp{\left(-\frac{x^2}{4Dt}\right)} $, then one may show that 
the van Hove function will have a long tail and depending on the
distribution of the $p(D,t)$, the tail of the distribution can be
either exponential or Gaussian \cite{12WKBG}. In general the
exponential tail has been reported \cite{07CBK} which, as mentioned in
\cite{12WKBG}, might be due to the small range of data. 

Now from the distribution of particle displacements one can
calculate the distribution of diffusivity just by defining $D \sim
\Delta r^2/\tau$, where $\Delta r^2$ is the square of the
displacements of particles over some time interval $\tau$. However, we feel
that the distribution of diffusivity calculated from the van Hove
functions is more generic and robust. To
substantiate our argument, lets take a system where diffusion
process is anisotropic $\it e.g.$ diffusion along microtubule in
biological context \cite{12WKBG} or diffusion of tagged particles
in nematic liquids \cite{sDhara12}. Now on top of diffusion, if
these molecules or tagged particles  experience an internal drift
force ($\it e.g.$ self propulsion) and one tries to
calculate the diffusivity from the displacement, then the
estimation of the diffusivity will be completely wrong. On the
contrary, the method mentioned here will be free from these
problems as the distribution of particle displacements along the
anisotropic axis will have a non-zero mean value and one can
easily decouple the over all drift of the system from the
diffusion process and can still extract the distribution of the
diffusivity without much difficulty. Although for isotropic systems
both the methods will give the same results, the latter will be
preferable in more general cases.  

The key assumption in the method described here is that the long
time dynamics of glass-forming liquids can be described by a
superposition of diffusive processes. This is justified by the
following argument: previous studies have shown that single
particle trajectories in glass forming liquids can be described as
short time vibrations inside cages with infrequent long time jumps
due to cage breaking and formation of new cages. Thus,
single-particle motion at long times and on a course grained
length scale of a cage can be considered as diffusion among
cages. We emphasize that particle displacements may be described
by diffusive processes only for time intervals of the order of the
time taken for the slope of the mean square displacement (MSD) to
reach its asymptotic value 1, {\it i.e.} at least of the order of
$\alpha$ relaxation times $\tau_{\alpha}$ (see Sec.\ref{sec:def}). 
Thus the distribution $P(D,t)$ is meaningful only for times 
greater than or equal to $\alpha$ relaxation times. We also note 
that the van Hove function is time dependent - having long time 
tail on the time scale of $\alpha$ relaxation times (when 
glass-forming liquids typically show maximum dynamical 
heterogeneity), but gradually becoming Gaussian in the limit 
of infinite time. Consequently the distribution of diffusivity 
must also be time dependent : $P(D,t)$ approaches a $\delta$ 
function asymptotically.  

Our main aim in this study is to calculate \emph {directly} this
distribution of the diffusivity from the simulation data and thus
understand the nature of DH. From the previous discussion, one may
expect that in deeply supercooled liquids the distribution of
diffusivity will be bimodal in general. Our results unambiguously
show that the distribution of diffusivity indeed becomes
bimodal below some temperature. However, the bimodal nature of the
distribution does not prove that particles are clustered together
to form ``solid like'' and ``fluid-like'' regions. Hence
bimodality alone is not enough to justify the picture of
supercooled liquids being a sparse mixture of ``fluid like'' and
``solid like'' regions. To test the existence of such clusters, we
employ another kind of numerical experiment, where we probe the
effect of random pinning on the system dynamics.   

Recent studies \cite{12KLP,13KP, 03Kim, 12BK, 13KB,
  cammarotaPinning, 12CBEPL, 13CGB, 11KMS,
  krakoviackPinning, szamelPinning} on dynamics of supercooled
liquids in the presence of quenched disorder have shaded some
interesting lights on the puzzles of glass transition including
possible existence of ideal glass transition with increasing
disorder strength \cite{cammarotaPinning}. In most of the cases
the quenched disorder is introduced by randomly freezing (pinning)
some fraction of the particles in the system and it was found that
relaxation time $\tau_{\alpha}$ increases drastically with
increasing pinning concentration. The effect is so dramatic that
in some simulation studies and subsequent mean-field and
renormalization group analyses it was suggested that one can reach
ideal glass state by simply increasing pinning fraction
\cite{12KLP,12BK,cammarotaPinning}. These studies have opened up a
new avenue for exploring and understanding the puzzles of glass
transition from a completely different perspective. One of the
main findings of our study along these directions is the dramatic
effect of random pinning on the SE break down. We show that the
degree of SE breakdown can be very efficiently tuned by
introducing random pinning in these model systems. Finally we
argue from these observed effects of random pins that slow and
fast particles indeed cluster to form ``solid like'' and ``fluid
like'' regions in the system over the time scale of relaxation
time, $\tau_{\alpha}$. 

%The phenomenological explanation of decoupling does not tell
%anything about the nature and origin of the
%heterogeneity. Microscopic theories {\it e.g.} mode coupling
%theory (MCT) \cite{pap:Biroli-Bouchaud}, random first-order
%transition (RFOT) theory \cite{pap:SE-Xia-Wolynes,
%pap:Frag-Xia-Wolynes, pap:SE-Lubchenko-Wolynes}, shear
%transformation zone (STZ) theory \cite{pap:SE-Langer}, dynamical
%facilitation \cite{pap:SE-Jung-Garrahan-Chandler} and the
%obstruction model \cite{pap:SE-Douglas-Leporini} do not mutually
%agree  on the origin and the nature of heterogeneous
%dynamics. Besides, the decoupling of $D$ and $\tau_{\alpha}$ may
%also be explained in terms of a growing length scale
%\cite{08Chong,chong} which does not directly require a
%distribution of diffusivity. 
The paper is organised as follows : first we specify the
simulation details and define the relevant quantities. Then we
briefly explain the method used to extract the distribution of
diffusivity and relaxation time. This methodology is then applied
to a model supercooled liquid to understand the origin of 
dynamic heterogeneity and Stokes-Einstein breakdown. In the 
end we used random pinning geometry to further strengthen 
our conclusions reached from the analysis of distribution of 
diffusivity and relaxation time.

\section{Simulation Details} \label{sec:simu}

In the present study, we analyzed the Kob-Andersen binary mixture
\cite{95KA} as a prototype glass-forming liquid. We performed NVT
MD simulations using periodic boundary conditions at the constant
number density $N/V = 1.2$, where $N = 1024$ was the system size and
$V$ was the volume of the system. The units of length, energy and time
were same as in Ref. \cite{95KA}. Integration time steps were $\delta
t = 0.005$ (in reduced units).  At each state point, different
quantities were averaged over $20$ different initial conditions, each
run being at least 100 $\alpha$ relaxation times (defined in
Sec. \ref{sec:def}) long.

\section{Definitions}\label{sec:def}

\paragraph{van Hove function :} 
The van Hove function in one dimension is defined as $G(x,t) =
\left \langle \frac{1}{N} \sum_{i=1}^N \sum_{j=1}^N \delta \left(
    x - x_j (t) + x_i (0) \right) \right \rangle$, where $x_j (t)$
is the $x$ coordinate of the position vector $\vec{r}_j (t)$ of
the $j$th particle at time $t$ and $<.>$ implies averaging over
time origins. In the present study,  we compute the self part of
the van Hove function (taking all particles) defined as

\begin{equation}
G_s (x,t) = \left \langle \frac{1}{N} 
  \sum_{i=1}^N \delta \left( x - x_i (t) + x_i (0)
  \right) \right \rangle
\end{equation}

\paragraph{Diffusivity ($D_\infty$) :} 
The mean squared displacement (MSD) is computed from $\left\langle
  \Delta r^2 (t) \right\rangle = \left\langle \sum_{i=1}^N
  \left[\vec{r}_i (t) - \vec{r}_i(0) \right] ^2 \right\rangle$
(averaging over both the species). We denote the diffusivity
computed from the long time limit of the MSD using the Einstein
relation as $D_\infty$ :

\begin{equation}
\lim_{t \rightarrow \infty} \left\langle \Delta r^2 (t)
\right\rangle = 6 D_\infty t
\label{eqn:D}
\end{equation}

\paragraph{Diffusivities for ``solid-like'' ($D_s$) and
  ``fluid-like'' ($D_l$) particles :}
In the present study we identify subsets of particles as less
mobile or ``solid-like'' or slow and more mobile or ``fluid-like''
or fast particles based on a distribution of diffusivity (see Sec
\ref{sec:intro} and \ref{sec:met} for details). At low
temperatures the distribution is multimodal and the peak positions
are used for estimating ``solid-like'' and ``fluid-like''
diffusivities. At a given temperature, a ``fluid-like''
diffusivity $D_l$ is estimated from the positions of the peak at
higher diffusivity.  Similarly a ``solid-like'' diffusivity $D_s$
is estimated from the position of the peak at lower
diffusivity. At the lowest three temperatures, where additional
shoulders appear, the position of the dominant peak at lower
diffusivity is taken to estimate $D_s$.
\vspace{2mm}

\paragraph{$\tau_m$, $\tau_l$ :} 
From the MSD, we compute two time scales $\tau_m$ and $\tau_l$
defined as the times when MSD at a given temperature respectively
becomes 0.50 and 1.00 (in units of squared particle diameter
$\sigma_{AA}^2$).

\begin{eqnarray}
\left\langle \Delta r^2 (\tau_m)
\right\rangle / \sigma^2_{AA}= 0.50 \nonumber\\
\left\langle \Delta r^2 (\tau_l)
\right\rangle / \sigma^2_{AA}= 1.00
\label{eqn:tmtl}
\end{eqnarray}

\paragraph{Relaxation time, $\tau_\alpha$ :} 
$\alpha$ relaxation times are computed from the decay of the
overlap function. The overlap function $Q(t)$ is a normalized two
point correlation function defined as \cite{12SKDS}

\begin{equation}
Q(t) = \left < \frac{1}{N}\sum_{i=1}^{N} w(|\vec{r}_i(t) - \vec{r}_i(0)|) \right >
\label{eqn:qt}
\end{equation}
with $w(x) = 1.0$ for $x< 0.30$ and zero otherwise. The summation
is over all particles except for system with frozen particles,
where the summation is only over the mobile particles and $N$ is
the number of such particles. The $\langle \cdots \rangle$ indicates the
averaging over time origin and different statistically independent
simulation runs. Then $\tau_\alpha$ is defined as  

\begin{equation}
Q(\tau_{\alpha}) = 1/e
\label{eqn:talpha}
\end{equation}

\paragraph{Relaxation times for ``solid-like'' ($\tau_s$)
  and ``fluid-like'' ($\tau_l$) particles:} 
To calculate the $\alpha$ relaxation time of the ``solid-like'' or
slow and ``fluid-like'' or fast particles, a suitable cut off is
defined from the distribution of the diffusivity calculated at
$\tau_m$. We choose the minimum of the distribution which appears
at $D\tau_m \sim 0.060$ as the obvious cut off to define the slow
and fast particles. A particle is defined to be slow if its
squared displacement is less than or equal to $\Delta r^2 =
6D\tau_m \sim 0.360$ in time $t = \tau_m$, otherwise we define it
as a fast particle. Once a set of slow and fast particles are
defined for one time origin, we calculate the overlap function for
slow $[Q_s(t)]$ and fast $[Q_l(t)]$ particles as

\begin{eqnarray}
Q_s(t) = \left< \frac{1}{N_s}\sum_{i=1}^{N_s} w(|\vec{r}_i(t) -
  \vec{r}_i(0)|)\right> \nonumber\\ 
Q_l(t) = \left< \frac{1}{N_l}\sum_{i=1}^{N_l} w(|\vec{r}_i(t) -
  \vec{r}_i(0)|) \right>,
\label{eqn:qtslowFast}
\end{eqnarray}
where $N_s$ is the number of slow particles and $N_l = N - N_s$ is
the number of fast particles. Notice that the number of slow and
fast particles changes with different time origins and we then
average all these correlation functions calculated at different
time origins to get the average correlation function. The
relaxation time $\tau_s$ or $\tau_l$ is then calculated as
$Q_s(\tau_s) = 1/e$ and $Q_l(\tau_l) = 1/e$.

\paragraph{Stokes Einstein (SE) relation :} 
The Stokes Einstein relation connects the translational
diffusivity  $D_B$ of a probe Brownian particle inside a viscous
liquid to the shear viscosity $\eta$ of the liquid : $D_B = c
\frac{k_B T}{\eta}$, where $c$ is a constant whose value depends
on the particle details and the boundary conditions and $T$ is
temperature. In the present study, as often done in the
literature, the diffusivity of the probe particle is replaced by
the self diffusivity ($D_\infty)$ of the liquid and the shear
viscosity is replaced by the $\alpha$ relaxation time
$\tau_\alpha$. Thus, we consider the following relation as the
Stokes Einstein relation :

\begin{equation}
D_\infty = c \frac{k_B T}{\tau_\alpha}
\end{equation}

\paragraph{SE violation parameters :}
In many glass-forming liquids, the SE relation breaks down at low
temperatures. The breakdown is characterized by a SE violation
parameter $\theta (T)$ defined as :

\begin{equation}
\theta (T) = \frac{D_\infty \tau_\alpha}{T} 
\end{equation}

In addition, we compute SE violation parameters with the
``solid-like'' and ``liquid-like'' diffusivities as : 

\begin{eqnarray}
\theta_s (T) = \frac{D_s \tau_s}{T}  \nonumber\\
\theta_l (T) = \frac{D_l \tau_l}{T}.
\end{eqnarray}

\paragraph{Fractional SE relation and the breakdown exponent $w$
  :} At low temperatures, where the SE relation $D_\infty \propto
\tau_\alpha^{-1}$ breaks down, many glass-forming liquids obey a
fractional SE relation \cite{SEB06BPS,pap:SE-Douglas-Leporini} : 

\begin{equation}
D_\infty \propto \tau_\alpha^{-1 + \omega}
\end{equation}
where $\omega$ is denoted as the SE breakdown exponent. If
$\omega=0$, the SE relation holds. In general,  $0 \leq \omega
\leq 1$.

\section{Distribution of Diffusivity}\label{sec:met}

Lets start with briefly describing the method used to extract the
distribution of diffusivity directly from the self van Hove
correlation function $G_s(x,\tau)$ using the iterative algorithm
suggested in \cite{74Lucy} and recently used in \cite{12WKBG} for
diffusion processes in biological systems. We assume that particle
displacements are caused by diffusion processes and there is a
distribution of local diffusivity $p(D,\tau)$. Then we can express
$G_s(x,\tau)$ in terms of $p(D,\tau)$ using
Eq.\ref{vanHoveEq}. Now given the $G_s(x,\tau)$, we calculate the
distribution  $p(D,\tau)$ following \cite{74Lucy} as 

\begin{equation}
p^{n+1}(D,\tau) = p^{n}(D,\tau)\int_{-\infty}^{\infty}
\frac{G_s(x,\tau)}{G^n_s(x,\tau)}g(x|D) dx,
\end{equation}
where $p^n(D,\tau)$ is the estimate of $p(D,\tau)$ in the $n^{th}$
iteration with $p^0(D) = (1/D_{avg})\exp(-D/D_{avg})$ and
\begin{equation}
G^n_s(x,\tau) = \int_0^{D_0} \, p^n(D,\tau)\, g(x|D) \, dD.
\end{equation}
We choose $D\tau$ as our variable because in the studied
temperature range $D$ changes by several orders of magnitude
whereas $D \tau$ changes relatively modestly. Hence 
\begin{equation}
P^{n+1}(D\tau, \tau) = P^{n}(D\tau, \tau)\int_{-\infty}^{\infty}
\frac{G_s(x,\tau)}{G^n_s(x,\tau)}g(x|D) dx,
\label{eqn:LucyDtau}
\end{equation}
where $p(D,\tau)dD = P(D\tau,\tau)d(D\tau)$. 

\begin{figure}[!h]
\includegraphics[scale = 0.420]{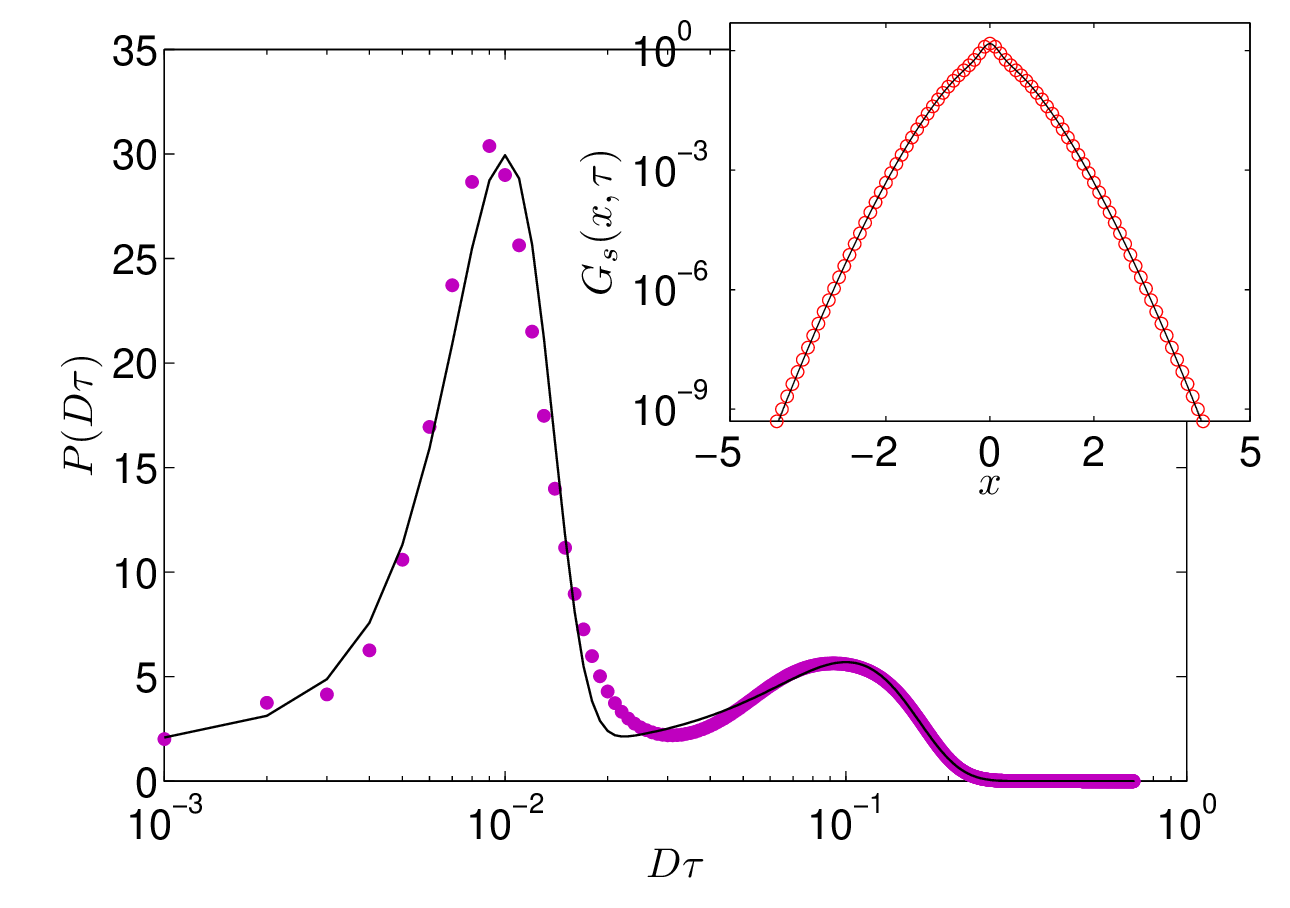}
\caption{The solid line shows the results of the iterative scheme
used to calculate the distribution of diffusivity (see text for
details) along with the original distribution shown as
symbols. The agreement is really encouraging and inset shows the
corresponding comparison for the van Hove correlation
functions.} 
\label{probDistToy}
\end{figure}

We tested this iterative scheme first on a toy model to check its
convergence. We started with a toy distribution defined
as $P(x)\equiv A \exp\left[-(x-\mu_1)^2 / 2 \sigma_1^2\right] + B
\exp\left[-(x-\mu_2)^2 / 2 \sigma_2^2\right]$ with the parameters
$A,B, \mu_1, \mu_2, \sigma_1, \sigma_2$ chosen by hand. Then we
calculated the toy van Hove correlation function $G_s(x,\tau)$
using Eq.\ref{vanHoveEq}, and used this as input to recalculate
the probability distribution $P(D\tau)$ using
Eq. \ref{eqn:LucyDtau}. In Fig. \ref{probDistToy}, we have 
compared the calculated distribution $P(D\tau)$ with the exact
distribution. The agreement is really good with moderate number of
iterations. In the inset of Fig.\ref{probDistToy}, we have
compared the van Hove function obtained from the converged
distribution of diffusivity with the exact one. Here also the
agreement is near perfect. Note that the iterative scheme does not
depend at all on the initial guess distribution as long as it is
non-negative and normalizable. 

\section{Distribution of Relaxation time}\label{tauMet}
According to the criterion mentioned in the definition section, 
we first define slow and fast particles and then calculate their
respective overlap correlation functions $Q_s(t)$ and
$Q_l(t)$. The distribution functions are then calculated from
these overlap functions using the following integral relation,

\begin{equation}
e^{-(t/\tau_{\alpha})^{\beta}} = \int_{0}^{\infty} e^{-t/\tau} P(\tau) d\tau.
\label{tauDist}
\end{equation}
where $\beta$ is the stretching parameter and also known as $KWW$
exponent in glass literature. Since the long time part of the
overlap function $Q(t)$ can be very well described by stretched
exponential form, one can try to extract the underlying
distribution of relaxation time $P(\tau)$ by inverting the above
equation. Analytical solution for this method is not directly
available but some work along this line in \cite{alvarez} provides
some useful insights and highlights difficulties associated with
this problem. Here we have extracted this distribution within the
Gaussian approximation by optimizing the following cost function

\begin{equation}
\chi^2 = \frac{1}{n}\sum_{i=1}^{n} \left[ Q_{\nu}(t_i) - 
\int_{0}^{\infty} e^{-t_i/\tau} P(\tau) d\tau \right]^2
\label{costFn}
\end{equation}
with respect to the parameters of the $P(\tau)$. $\nu$ denotes slow 
(s) or fast (l) particles index. We choose the
following functional form for $P(\tau)$

\begin{equation}
P(\tau) = \frac{1}{\tau \sqrt{2\pi}\sigma} e^{-\frac{(\log(\tau) -
    \log(\tau_0))^2}{2\sigma^2}}
\end{equation}
such that $\int_{0}^{\infty}P(\tau)d\tau = 1.0$. This is partly 
motivated by Ref.\cite{alvarez}, where the formal expression of 
the distribution is written in terms of logarithm of the variable. 
So within Gaussian approximation, it becomes log-normal in $\tau$.
\begin{figure}[!h]
\includegraphics[scale = 0.350]{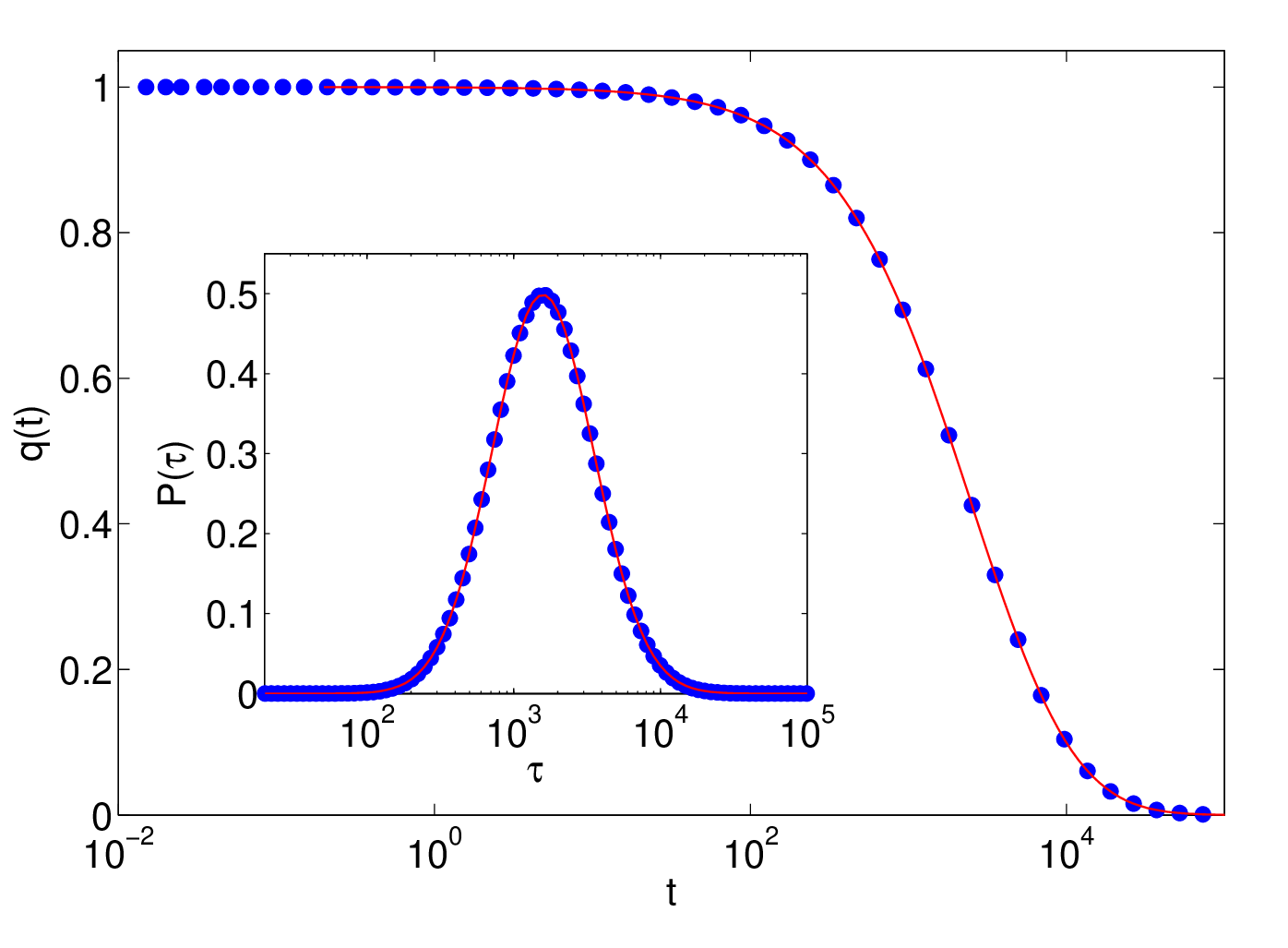}
\caption{Plot of the stretched exponential function $q(t)$ (filled
  circles) as a function of time and the line is the corresponding
  output from the optimization of cost function in
  Eq.\ref{costFn}. Inset: the comparison of the distribution of
  $\tau$ (filled circle) with the solution obtained from the
  optimization procedure (see text for details). The convergence
  of this optimization is very quick and does not depend on the
  initial parameter value.}
\label{tauToy}
\end{figure}

To check the convergence of this method to the correct solution, we
first tested it on a toy problem where we have generated a
stretched exponential function $q(t)$ by using Eq.\ref{tauDist}
with a set of chosen values of $\tau_0$ and $\sigma$ and then
performed the optimization of the cost function in
Eq.\ref{costFn}. In Fig.\ref{tauToy}, we show the convergence of
this optimization procedure for the toy function. The convergence
is quick and does not depend on the initial guess values of the
parameters, $\tau_0$ and $\sigma$. In the result section we then
use this optimization procedure to extract the distribution of
relaxation time associated with the slow and fast particles {\it
  i.e.} $P(\tau_s)$ and $P(\tau_l)$ (see Results section for 
further details).

\begin{figure}[!h]
\includegraphics[scale = 0.25]{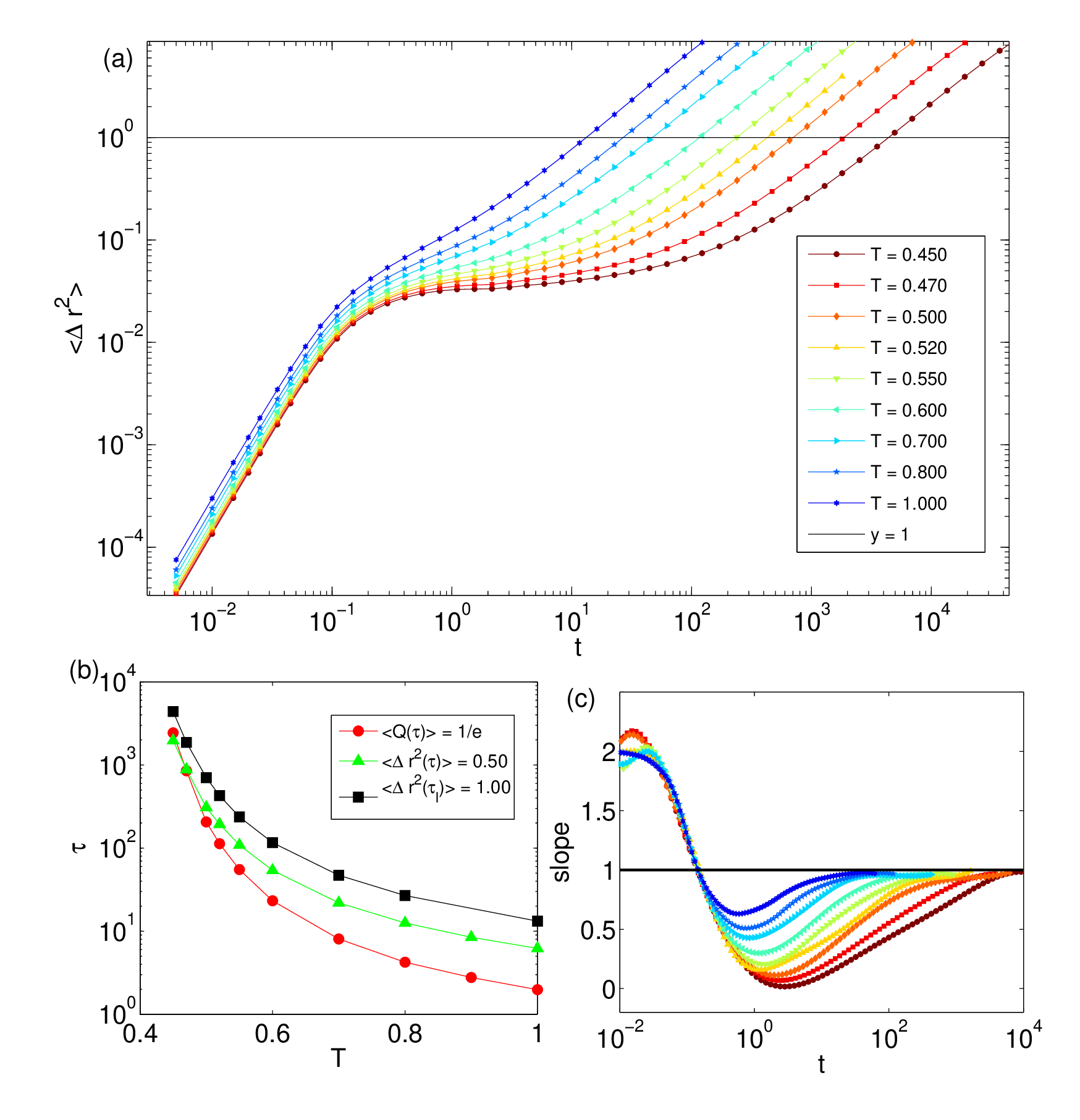}
\caption{(a) Mean Squared displacement (MSD) for different
  temperature with the horizontal line indicating the time where
  mean squared displacement reaches $1.00$ in reduced units for
  different temperatures. The corresponding time is denoted as
  $\tau_l$ here. Correspondingly $\tau_m$ is defined as the time
  where the MSD goes to $0.50$ in reduced units. (b) Comparison of
  the $\alpha$-relaxation time $\tau_{\alpha}$ with $\tau_m$ and
  $\tau_l$ (see text for definitions). (c) Time dependence of the
  slope $\frac{d \ln <\Delta r^2(t)>}{d \ln t}$ of the MSD at
  different temperatures. } 
\label{msd2}
\end{figure}

\begin{figure}[!h]
\includegraphics[scale = 0.33]{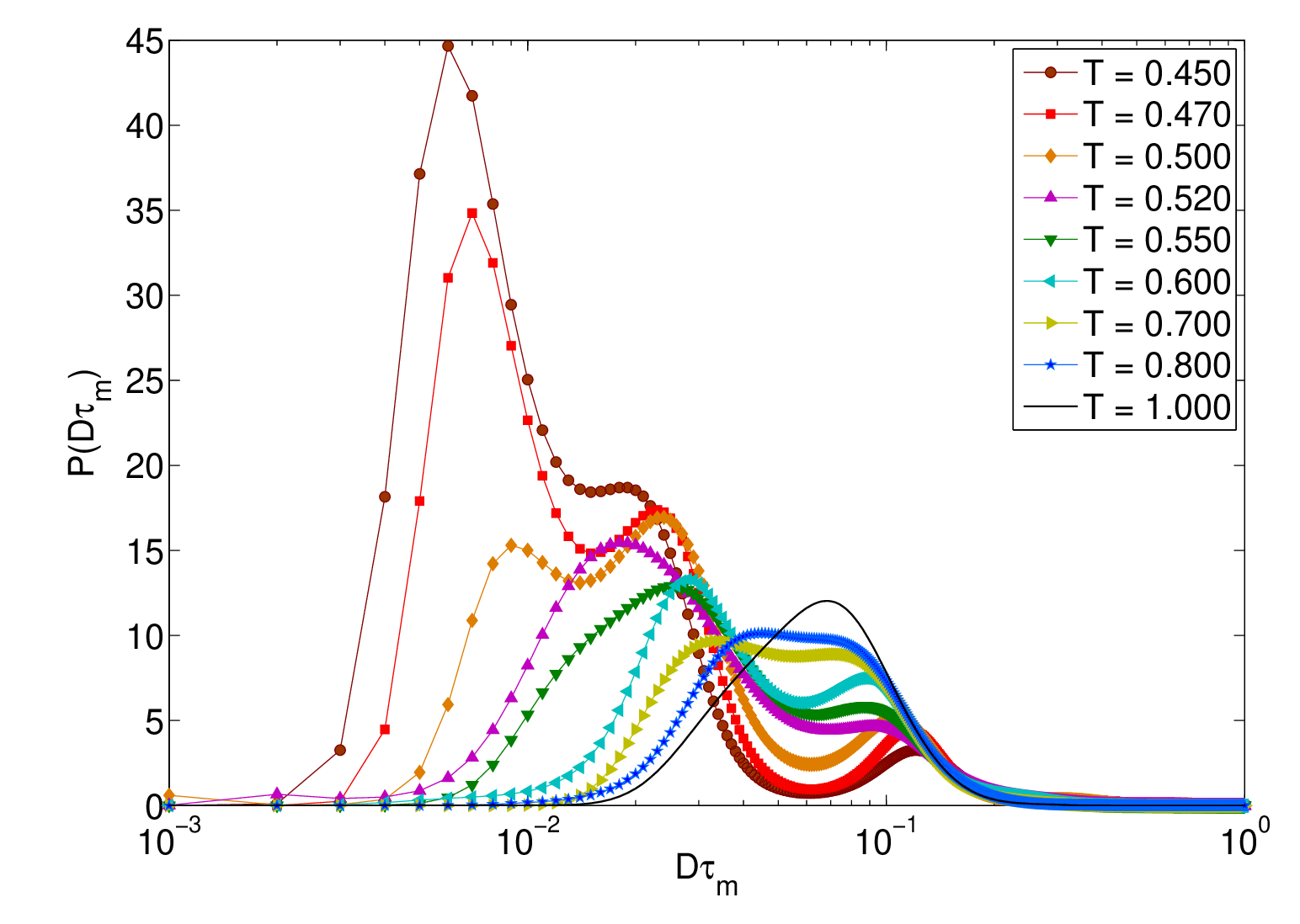}
\caption{Calculated distribution of diffusivity at times $\tau_m$
for different temperatures. Notice the appearance of the
bimodality in the distribution just below the onset temperature $T
= 1.00$. Clear bimodal distribution in the supercooled regime
confirms that there are two different types of particles in terms
of their mobility up to time scales of typical relaxation
times. The appearance of more peaks in the distribution at still
further lower temperature is really interesting, indicating
possibility of extremely slow to moderately slow to very fast
particles in very deep supercooled state.}
\label{diffDist}
\centerline{\includegraphics[scale = 0.400]{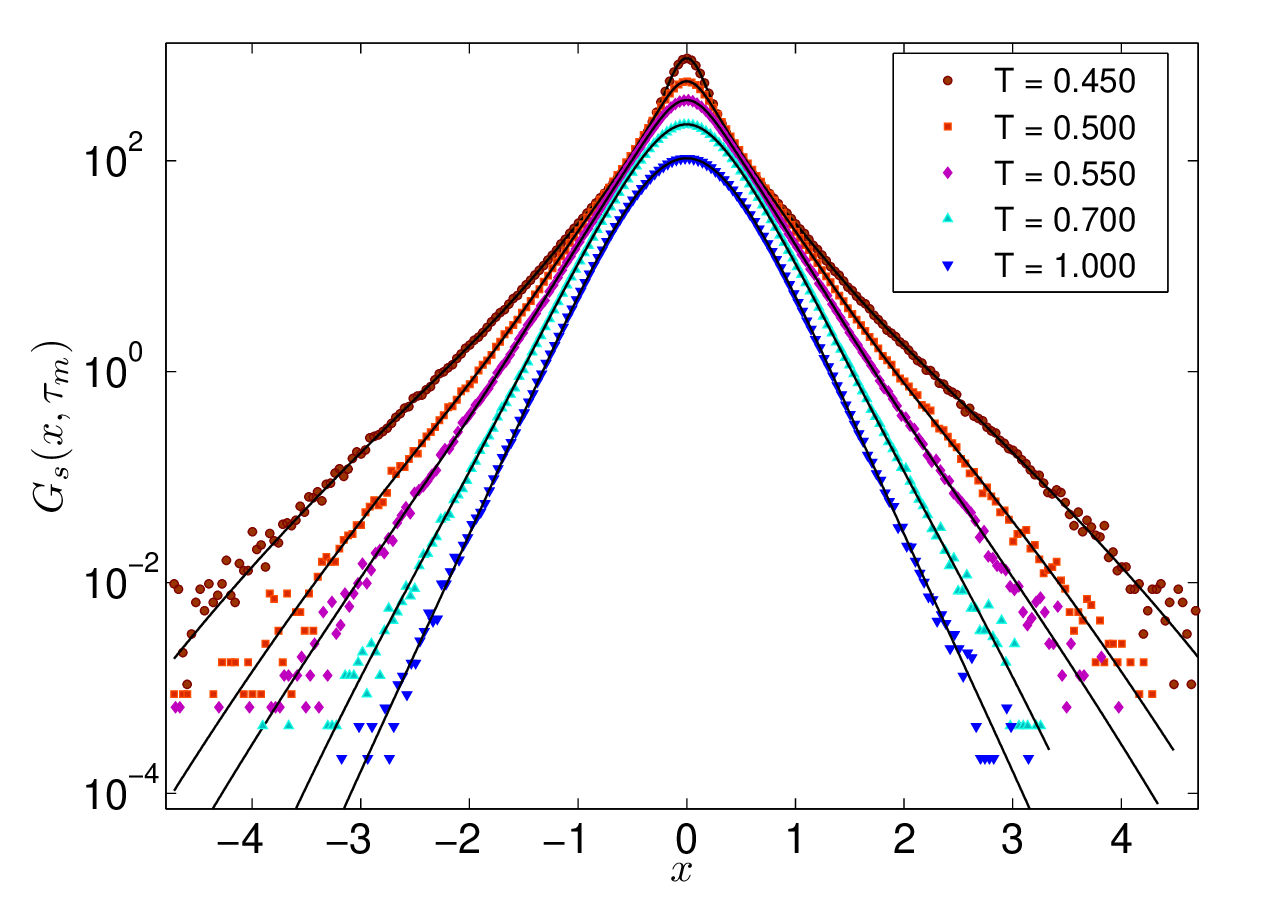}}
\caption{The van Hove correlation functions at times $\tau_m$
(solid line) as obtained from the iterative scheme (see text for
details) along with the simulation data (symbols). The curves are
shifted upward for clarity and to point out that the line goes
though the data points over the whole range of the data.}
\label{vanHoveFun}
\end{figure}

\section{Results}

\begin{figure}[!h]
\includegraphics[scale = 0.35]{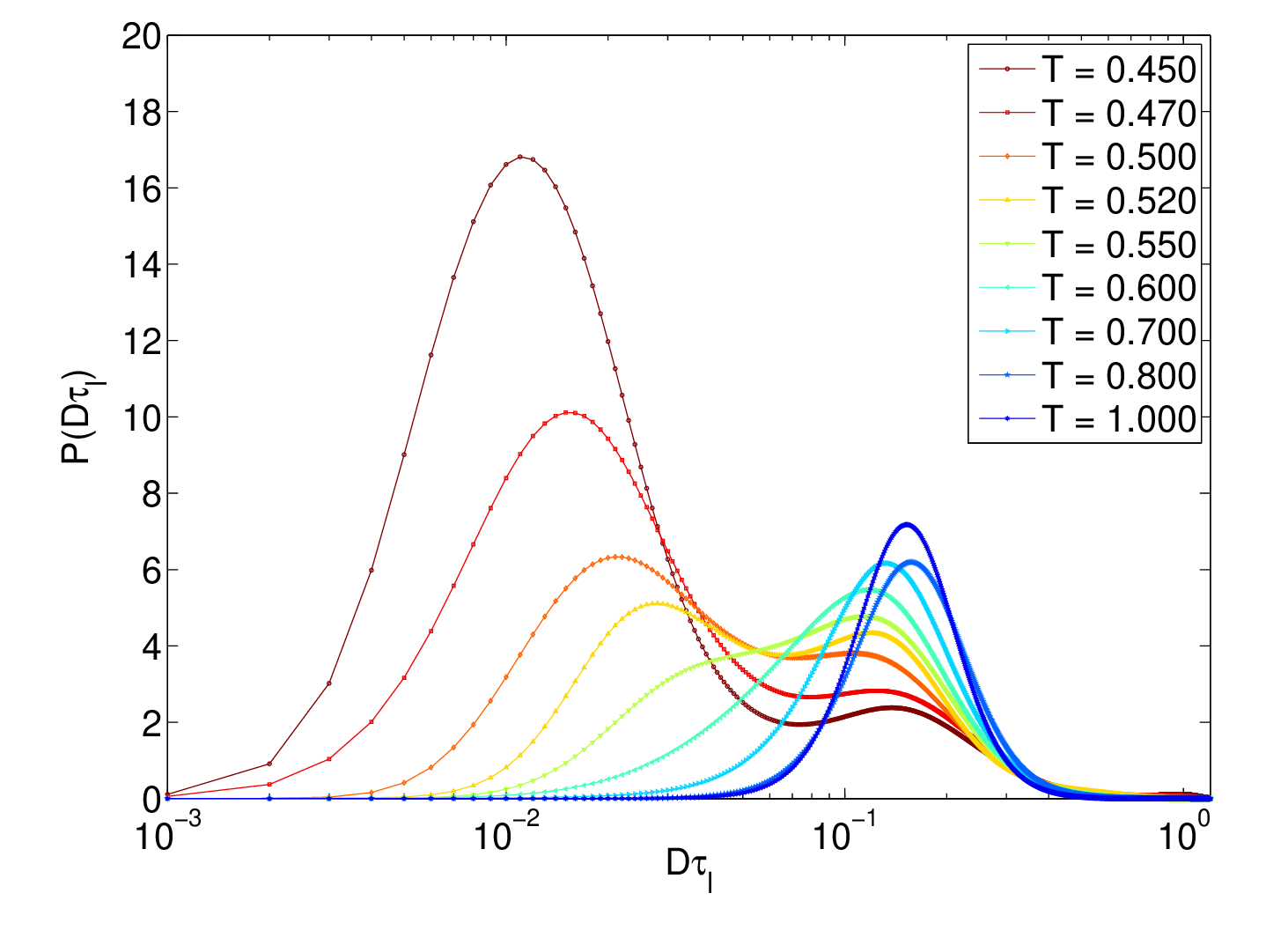}
\includegraphics[scale = 0.38]{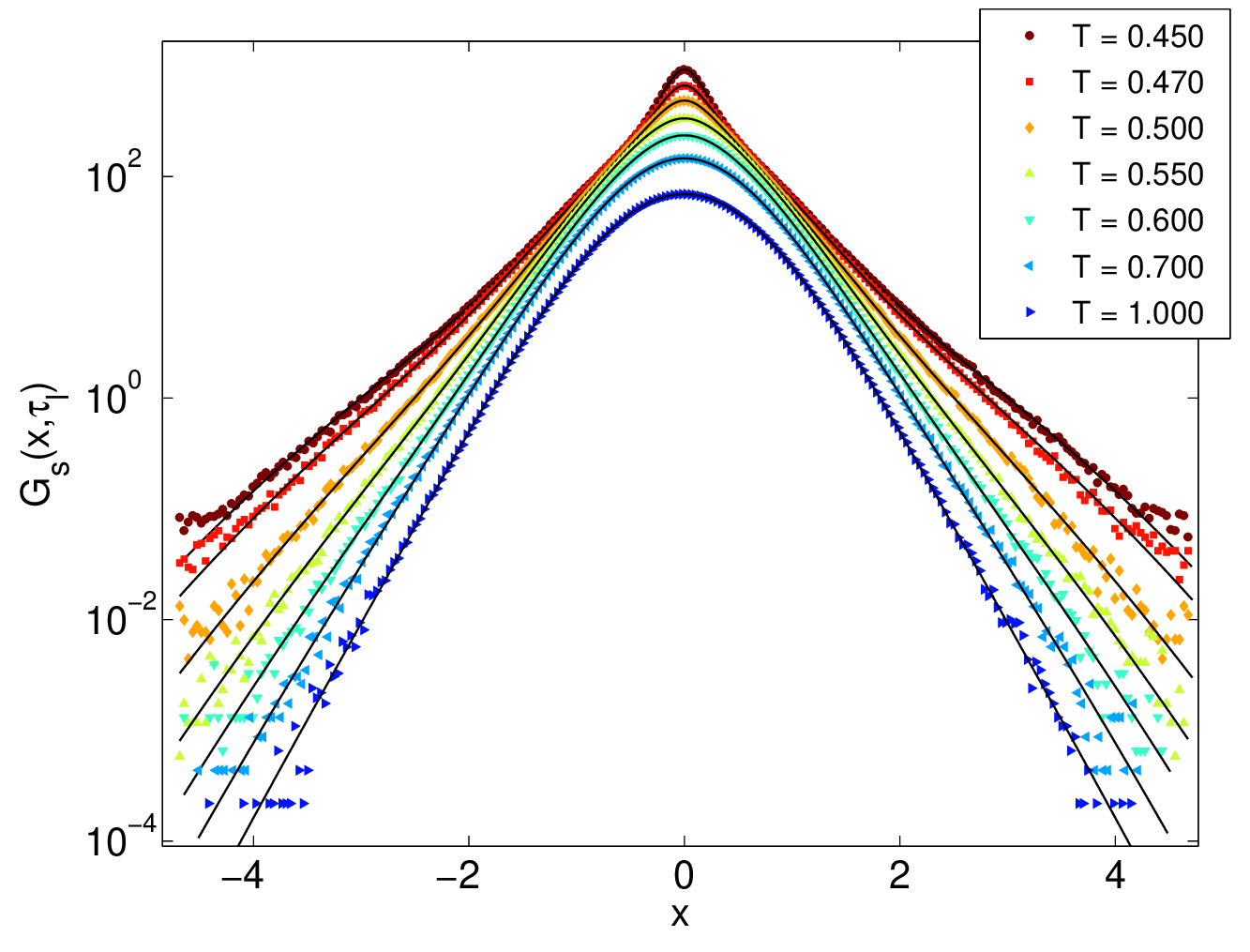}
\caption{Top: The distribution of diffusivity at times $\tau_l$
  for different temperatures. Even at $\tau_l$, the distribution
  is unimodal at high temperatures and bimodal at low
  temperatures. Bottom : Comparison of the van hove functions 
  at times $\tau_l$ at different temperatures, computed from the
  diffusivity distribution (solid lines) and directly measured
  (symbols). } 
\label{later}
\end{figure}

After establishing the rapid convergence of the iterative scheme
we tried to calculate the distribution of diffusivity for a model
glass forming liquid, the Kob-Andersen binary mixture \cite{95KA}.
We calculated the self van Hove functions at different
temperatures for the times $\tau_m$ and $\tau_l$ when the mean
square displacement (MSD) [see panel (a) of Fig.\ref{msd2}] 
becomes half and one respectively in reduced units. In the panel
(b) of Fig.\ref{msd2}, we have shown the temperature dependence of
these different time scales. It is important to mention that
$\tau_l$ is roughly the time where the logarithmic time derivative
of MSD with time $\frac{d \ln  \left\langle \Delta
    r^2(t)\right\rangle}{d \ln  t}$ becomes very close to $1.0$
[see panel (c) of Fig.\ref{msd2}]. This is the time when the
underlying relaxation process can be well approximated by
diffusion. We have repeated our calculation at two different time
scales just to make sure that the outcome is not an artifact of
doing the calculation at somewhat shorter time scale $\tau_m$
close to $\alpha$ relaxation times $\tau_{\alpha}$. Although
$\tau_m$ is of the order of $\tau_{\alpha}$, the temperature
dependence of $\tau_m$ is different from that of the
$\alpha$-relaxation time $\tau_{\alpha}$. 

At this point we would like to justify our choice of using
$\tau_m$ (or $\tau_l$ ) for analysis. We wanted to compare a
quantity (self van Hove function) which is a function of
displacement across different temperatures and a natural choice is
to have the mean squared displacement for these different
temperatures to be same. We can also take a fixed time (say
structural relaxation time or some multiple of it) and then
compare the function but our guess is that the qualitative result
will not change much. The temperature where the distribution of 
diffusivity calculated from self van Hove function starts to show
bimodality (discussed later) may change a bit depending on this
choice. 

\begin{figure}[!t]
\includegraphics[scale = 0.400]{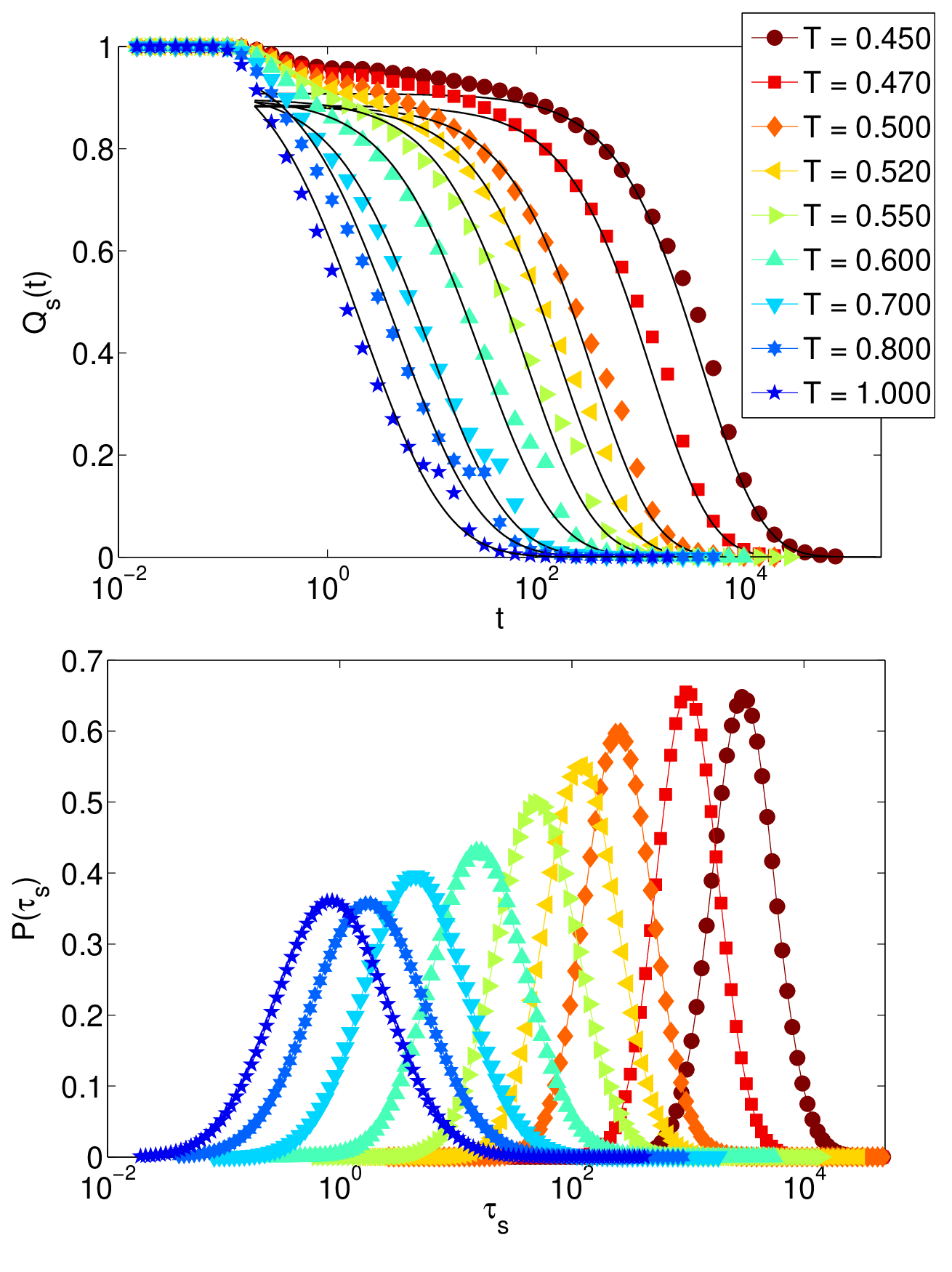}
\caption{Top Panel: Overlap correlation function $Q_s(t)$ for the
  slow particles for different temperatures (filled symbols) and
  the lines are the corresponding best fit stretched exponential
  function obtained by optimizing Eq.\ref{costFn} with the
  corresponding best estimation of the distribution functions (
  bottom panel).}
\label{slowFitting}
\end{figure}

To extract the distribution functions of diffusivity one needs to
supply somewhat smoothly averaged data of the van Hove functions
for the iterative scheme to converge rapidly. We fitted the
extreme tails of the calculated van Hove functions using
exponential functions as the tails are in general noisy and
difficult to average. We have shown the distributions
$P(D\tau_m,\tau_m)$ for different temperatures in
Fig.\ref{diffDist} and the van Hove functions calculated from
these distributions along with the simulation data in
Fig.\ref{vanHoveFun}. The agreement between the simulation data
and the calculated ones is indeed very good. Notice that tails of
these van Hove functions can not be completely described by a
single exponential function over the whole range at least for the
low temperature data ($T \le 0.50$). Rather they are better fitted
by two exponential functions. Similar analysis at a later time
$\tau_l$ shows no qualitative change in our results
(Fig.\ref{later}). Note that these results do not change with
different dynamics {\it e.g.} Brownian dynamics, as MSD from
molecular dynamics and from Brwonian dynamics become identical at
time scales of the order of $\tau_{\alpha}$.

\begin{figure}[!b]
\includegraphics[scale = 0.400]{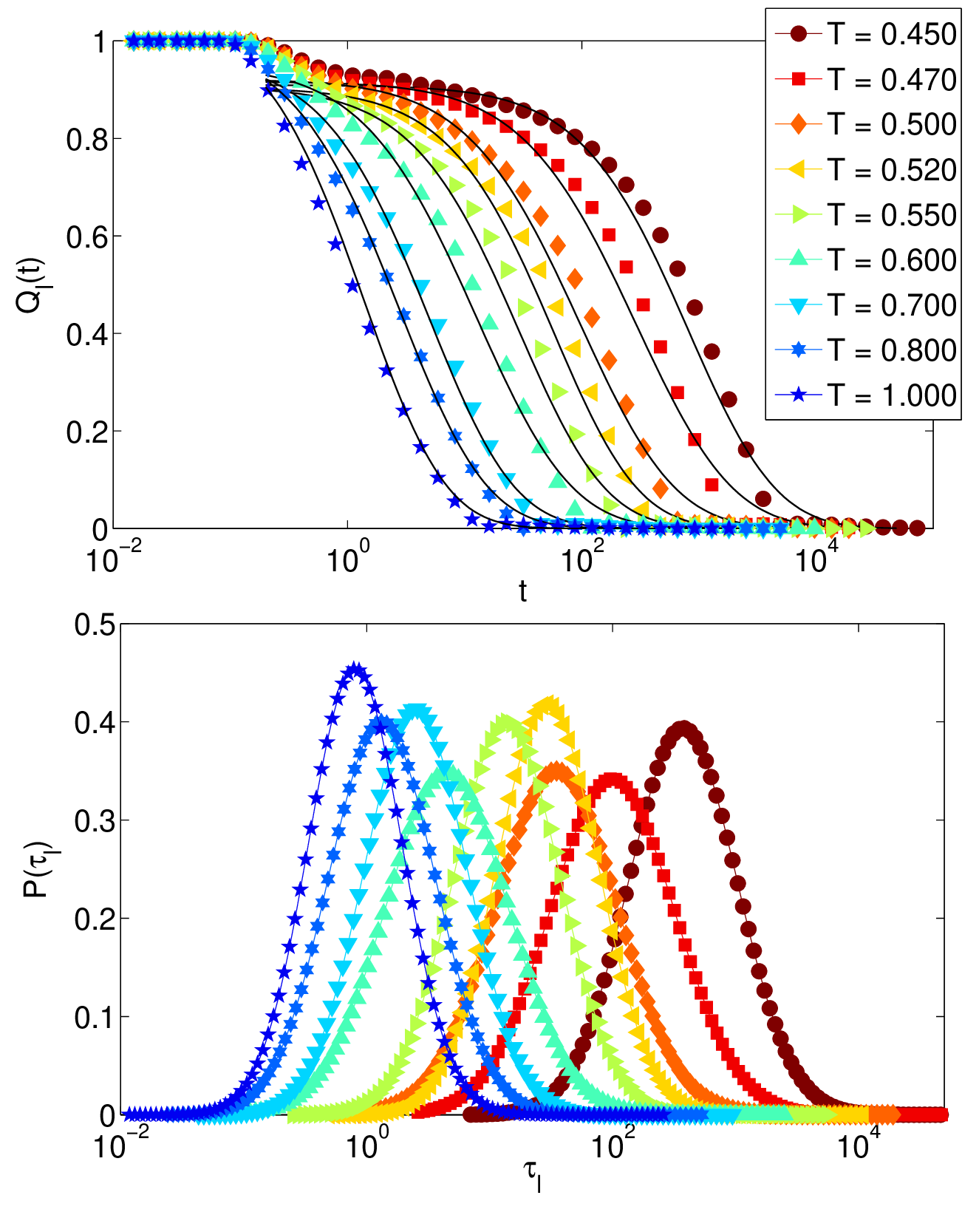}
\caption{Top Panel: Overlap correlation function $Q_l(t)$ for the
  fast particles for different temperatures (filled symbols) and
  the lines are the corresponding best fit stretched exponential
  function obtained by optimizing Eq.\ref{costFn} with the
  corresponding best estimation of the distribution functions (
  bottom panel).}
\label{fastFitting}
\end{figure}

Now looking at Fig.\ref{diffDist}, we can see that just below the
onset temperature ($T = 1.00$), the distribution starts to become
bimodal and two peaks clearly emerge at temperature around $T =
0.60$.  At further lower temperatures the distributions seem to
show existence of shoulders or another peak but the peak at large
diffusivity remains intact with decreasing peak height. Thus we
have clearly demonstrated that there are two types of particles in
the supercooled liquid on the time scale of the order of
$\tau_{\alpha}$. Notice that the width of the distribution
increases with decreasing temperature which indicates increase of
DH leading to stronger SE breakdown \cite{12SKDS}.

We have calculated the distribution of the relaxation times $\tau_l$
and $\tau_s$ using the optimization procedure described in
Sec.\ref{tauMet} from the respective overlap correlation functions
$Q_l(t)$ and $Q_s(t)$. We used data for $Q_s(t)$ or $Q_l(t)$ in
range $0<Q_s(t)<0.85$ and $0<Q_l(t)<0.85$ for the optimization as
short time part of the overlap function can not be represented by
stretched exponential form. In top panel of Fig.\ref{slowFitting}
we have shown the overlap  correlation function for the slow
particles for different temperatures and the corresponding best
fit stretched exponential functions obtained by optimizing the
cost function in Eq.\ref{costFn} with respect to the parameters of
the distribution function $P(\tau_s)$. The optimized distributions
themselves are shown in bottom panel of Fig.\ref{slowFitting}. One
can see that stretched exponential function obtained are quite
good fit to the $Q_s(t)$ data at higher temperature and seems to
become little bad at lower temperature but overall it is quite
good. It will be nice to be able to extract this distribution
without the Gaussian approximation and work along this line is in
progress. At this point we would like to point out that
distribution of relaxation time calculated in \cite{PH99} is
essentially same as that of the distribution of the diffusivity as
it was calculated by estimating the time taken by individual
particles to move a certain distance and this time is nothing but
diffusion time. In Fig.\ref{fastFitting}, we showed the results of
similar analyses done for the fast particles.

\begin{figure}[!b]
\centerline{\includegraphics[scale =0.380]{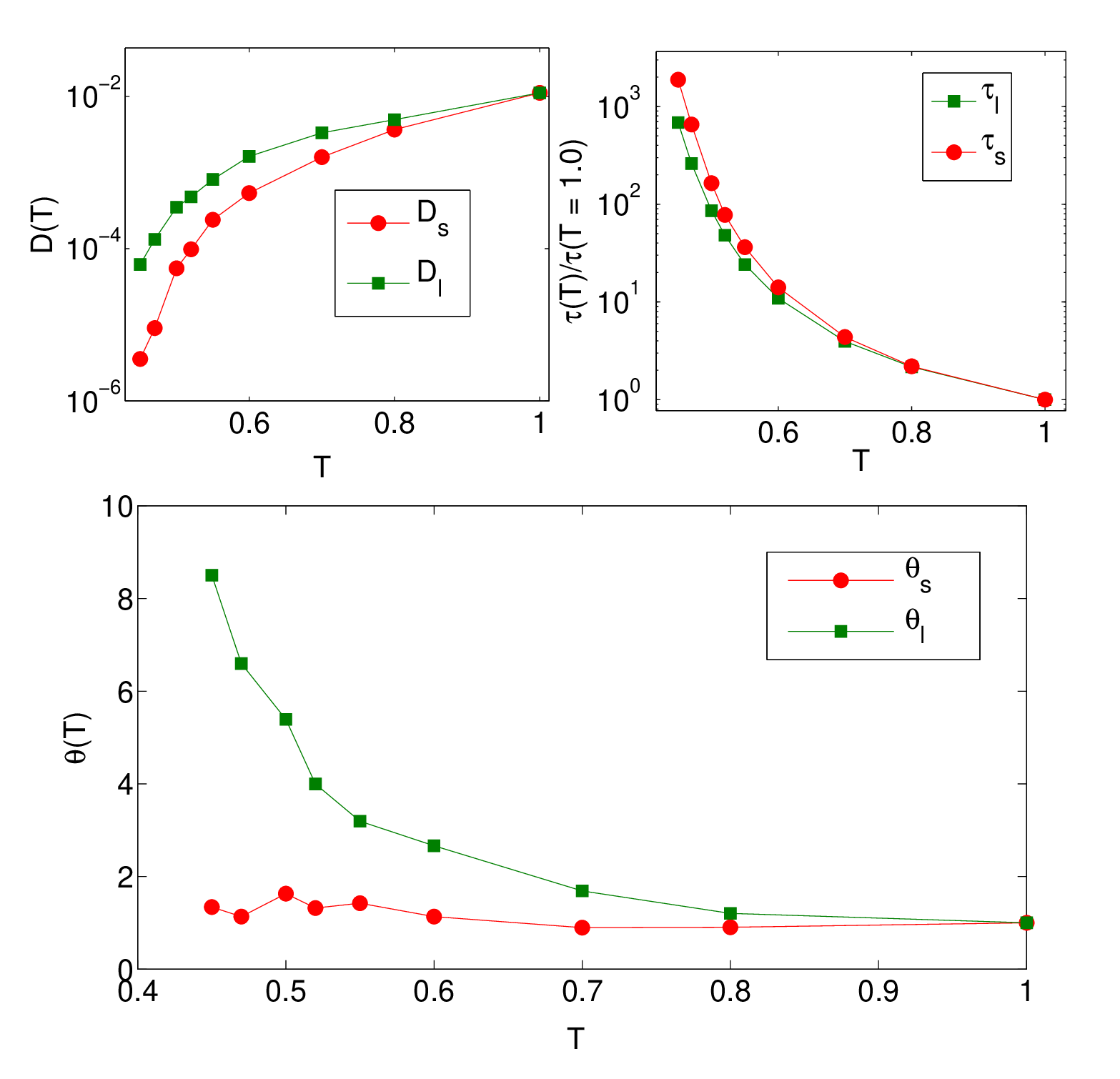}} 
\caption{Top left panel: The temperature dependence of the
  diffusivity associated with the solid like ($D_s$) and fluid
  like ($D_l$) particles. These values are calculated from the
  peak positions of the distributions in Fig.\ref{diffDist}. Top
  right panel: The temperature dependence of the relaxation times
  of slow ($\tau_s$) and fast ($\tau_l$) particles (see text for
  definitions). Bottom panel: The Stokes-Einstein violation
  parameters $\theta_s(T)=D_s\tau_s/T$ (circle) and $\theta_l (T)
  = D_l \tau_l / T$ (square) for the two types of particles. One
  sees clearly that solid like particles obey Stokes-Einstein
  relation to a reasonable accuracy over the whole temperature
  range and it is the fluid like particles which show strong
  Stokes-Einstein violation.}
\label{SEB}
\end{figure}
\begin{figure*}[!ht]
\includegraphics[scale = 0.350]{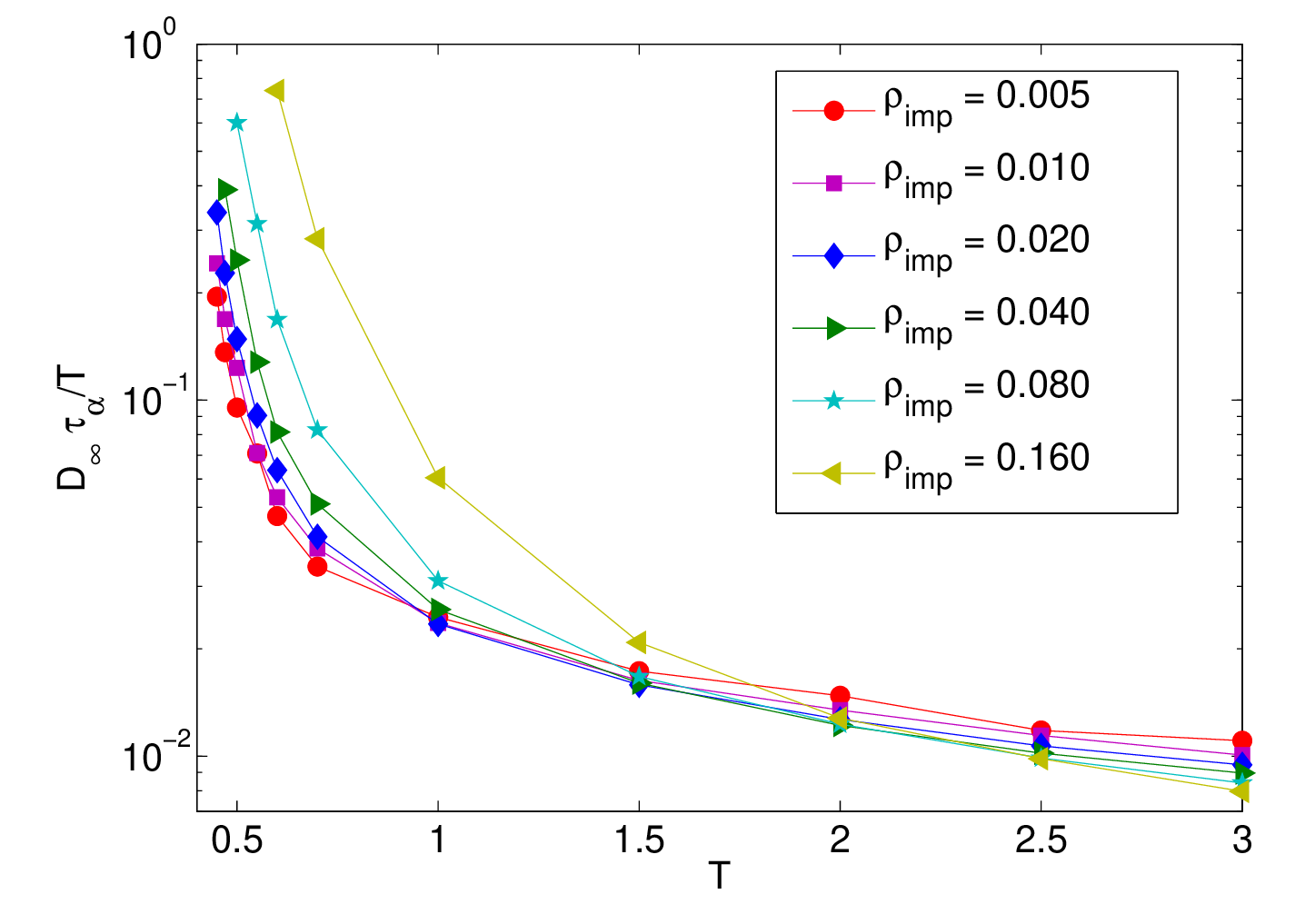}
\includegraphics[scale = 0.300]{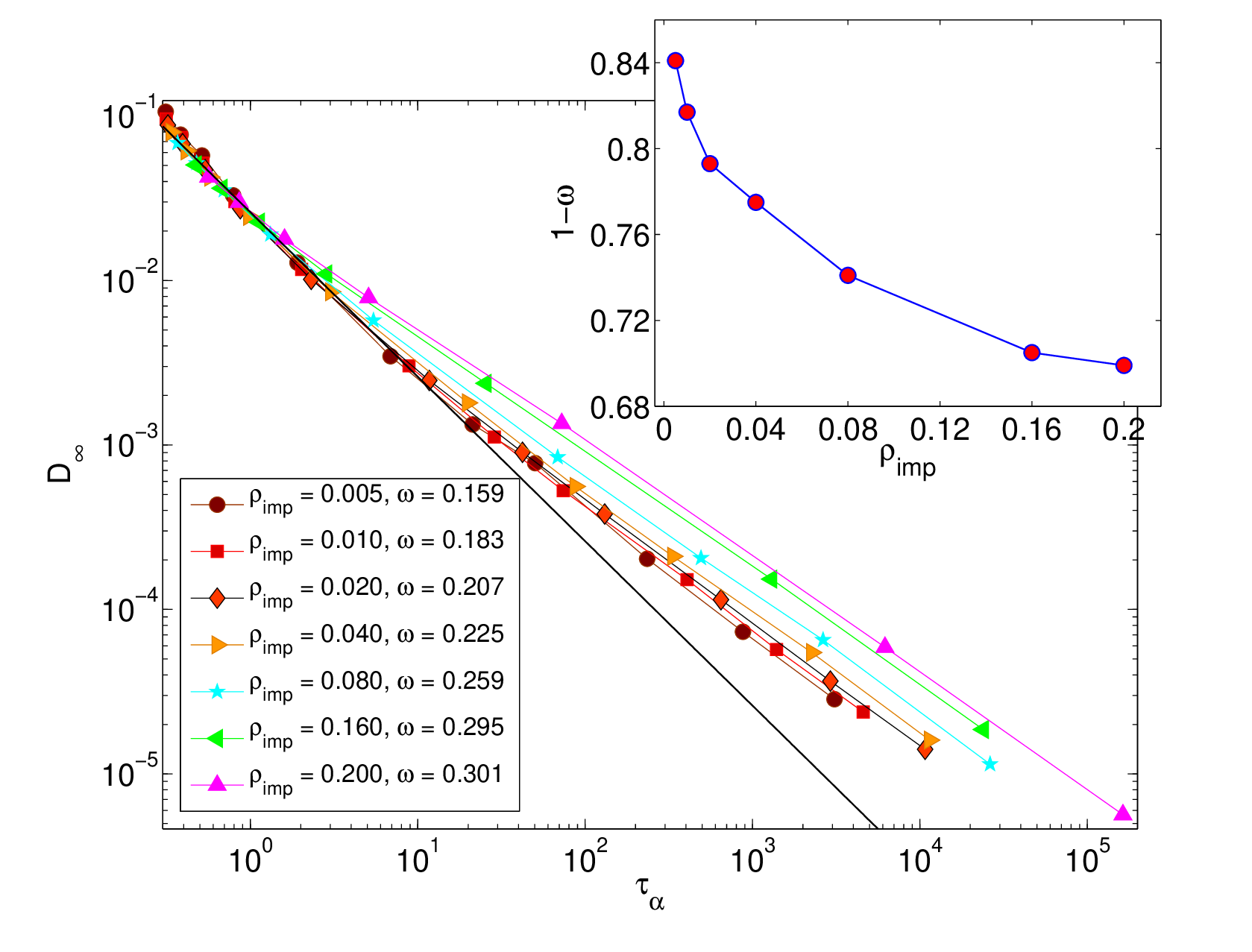}
\caption{Left Panel: Stokes-Einstein violation parameter
  $\theta(T)$ has been shown as a function of temperature for
  different density of the frozen particles $\rho_{imp}$. One can
  clearly see that the deviation from the Stokes-Einstein relation
  becomes stronger with increasing density of the frozen
  particles. Right Panel: Diffusivity ($D_\infty$) plotted as a
  function of relaxation time ($\tau_{\alpha}$) for different
  density of the frozen particles $\rho_{imp}$ in log-log to show
  the power law relationship between these quantities expected
  from the fractional Stokes-Einstein Relation $D_\infty \propto
  \tau_{\alpha}^{-1 + \omega}$, with $\omega \ge 0$. Notice that
  the exponent $\omega$ increase with increasing pinning density
  $\rho_{imp}$ indicating a stronger breakdown of the
  Stokes-Einstein Relation. The solid line has a slope equal to
  $-1.0$. Inset: The SE breakdown exponent as a function of
  $\rho_{imp}$. } 
\label{SEBRandomPinning}
\end{figure*}

After obtaining the distribution of diffusivity and relaxation
times for the slow and fast particles respectively, we now try to
understand the SE breakdown for both slow and fast particles
separately. In top left panel of Fig.\ref{SEB}, we have shown the
temperature dependence of diffusivities associated with the solid
like ($D_s$) and the fluid like ($D_l$) particles (defined in
Sec. \ref{sec:def}). The corresponding time scales $\tau_s$ and
$\tau_l$ are shown in the right panel of the same figure. In the
bottom panel of Fig.\ref{SEB}, we calculated the Stokes-Einstein
violation parameters $\theta_s(T)= D_s\tau_s/T$ and $\theta_l (T)
= D_l \tau_l / T$ for the two sets of particles. One clearly sees
that solid like particles obey the Stokes-Einstein relation over
the whole temperature range, whereas the fluid like particles show
strong SE violation leading to overall violation of the SE
relation in the liquid.

Note that in general one can think of the following scenarios 
which can lead to the SE breakdown in supercooled liquids:  
\begin{enumerate}
\item{Both slow and fast particles violate the SE relation
    simultaneously \cite{SEB06BPS}.} 
\item{Slow particles obey the SE relation and fast particles lead
    to violation.} 
\item{Both slow and fast particles obey SE relation separately but
    the combined effect leads to SE violation when one considers
    average relaxation time and the bulk diffusion coefficient.}  
\item{Slow particles violate but fast particles obey the SE
    relation.}
\end{enumerate}

So in this work we are able to rule out some of these
possibilities and show that the SE break down is caused by the
fast moving particles only. It should also be noted that we have
defined slow and fast particles using some cut off parameter and
the distributions are calculated based on this
definition. Although the cut off parameter chosen in this work is 
not completely adhoc, the results might depend on the specific
choice of this parameter. We believe that results will be fairly
insensitive to small variation in the value of the cut off
parameter and conclusion reached on the cause of the SE breakdown
will be same over that range. Thus a method to determine the
distribution of local relaxation times without invoking any
arbitrary cut off parameter will be very much desirable to remove
all these ambiguities in understanding the SE breakdown in
supercooled liquids.

After identifying two types of particles in the system on the time
scale of $\alpha$ relaxation time, we now turn to the question of
whether the ``solid-like'' particles form clusters. To answer
this, we performed simulations with some fraction $\rho_{imp}$ of
the particles randomly frozen in space and time and studied the
effect of this protocol on the dynamics of the system
\cite{12KLP,13KP, 03Kim, 12BK, 13KB, 11KMS}. If the ``solid like''
particles form clusters then if we \emph{randomly} freeze some
particles, there will always be instances where these frozen
particles are part of the ``solid like'' regions. In that case,
due to the frozen particles, the relaxation of these regions will
be hindered further and the relaxation time of the whole system
will increase dramatically with increasing density of these frozen
particles. However, these frozen particles will have very little
effect on diffusivity which is mainly governed by the ``fluid
like'' particles, so the diffusivity will not change dramatically.  
In this scenario, we expect to see an enhancement of the
Stokes-Einstein breakdown with increasing $\rho_{imp}$. On the
contrary if ``solid-like'' particles do not form clusters, then we
expect that increasing $\rho_{imp}$ will have little effect on the
relaxation dynamics of these regions  {\it i.e.} no enhancement of
the SE breakdown with $\rho_{imp}$.  In the left panel of 
Fig.\ref{SEBRandomPinning}, the temperature dependence of the SE
violation parameter $\theta(T) = D_\infty \tau_{\alpha}/T$ for
different pinning densities $\rho_{imp}$ shows dramatic deviation
of $\theta(T)$ from constancy with increasing
$\rho_{imp}$. Equivalently, by representing our data in $D_\infty$
{\it vs.} $\tau_{\alpha}$ (log-log) plot  in the right panel of 
Fig.\ref{SEBRandomPinning}, we see that in the low temperature
range the system obeys a \emph{fractional} Stokes-Einstein
relation $D \propto \tau_{\alpha}^{-1 + \omega}$, with $\omega \ge
0$, where the SE breakdown exponent $\omega$ increases
significantly with increasing $\rho_{imp}$. In the past, many
people have tried to tune the Stokes-Einstein breakdown exponent
to better understand the physics behind it,  either by tuning 
interaction potentials \cite{pap:SE-Affouard2009} or by changing
spatial dimensions \cite{DIM09ER,DIM10CIMM,DIM12CPZ} as one
expects mean field results of $\omega = 0$ to be exact at very
large spatial dimension. Our results provides yet another way of
tuning the SE breakdown exponent $\omega$ which we feel will have
interesting implications on the possible existence of ideal glass
transition with increasing pinning concentration as predicted in
\cite{cammarotaPinning}. 

\section{Conclusions}

To conclude, (a) we have directly calculated the distributions of
diffusivity for different temperatures for a model glass former in
the supercooled regime. This unambiguously shows that the state of
the system can be well described by a mixture of ``fluid like''
and ``solid like'' particles on the time scale of $\alpha$
relaxation time. This enabled us to extract the distributions
of relaxation times for slow and fast particles and helped us 
to discover that only fast particles are responsible for the 
Stokes-Einstein breakdown in bulk supercooled liquids; 
(b) we demonstrate, using a procedure (random
pinning) which does not involve any arbitrary cut-off, that the
``solid-like'' particles form clusters on the time scale of
$\alpha$ relaxation time;  and (c) finally we show that random
pinning can drastically enhance the decoupling between the
translational diffusion and the relaxation time, thereby providing
a new and easy way to tune the SE breakdown exponent.

The last result we believe is very significant in understanding
the origin of the SE breakdown in supercooled liquids. In
\cite{DIM12CPZ}, it is conjectured that local hopping,
facilitation and dynamical heterogeneity together play significant
roles in SE breakdown and the contribution of dynamic heterogeneity
might become smaller with increasing dimensionality but how other
contributions changes with dimensionality is not very clear. Now
random pinning might help us to resolve this issue. For
example it will be useful to understand whether the increase in
the SE breakdown exponent $\omega$ is related to the increase in
dynamic heterogeneity in the random pinning case also. It will 
also be important to study the SE breakdown by simultaneously
changing both dimensionality and fraction of randomly pinned
particles to pin point the correct reason behind the SE
breakdown. Finally it will be interesting to also extract the
length scale associated with the ``solid like'' and ``fluid like''
regions and compare that with the dynamic heterogeneity length
scale \cite{09KDS} and the static length scales
\cite{12KLP,SauTar10,BeKo_PS,HocRei12,12KP, 08BBCGV, 13BKP}. 

We would like to thank Srikanth Sastry, Chandan Dasgupta and
Saurish Chakrabarty for many useful discussions.

\end{document}